\documentclass[5p,twocolumn]{elsarticle}

\usepackage{graphicx}
\usepackage{dcolumn}
\usepackage{pifont}
\usepackage{bm}
\usepackage{multirow}
\usepackage{amsmath}
\usepackage{float}
\usepackage{txfonts}
\usepackage[version=3]{mhchem}

\journal{Elsevier}

\begin{document}
\title{Influence of water intercalation and hydration on chemical decomposition and ion transport in methylammonium lead halide perovskites}

\author[kimuniv-m,kimuniv-n]{Un-Gi Jong}
\author[kimuniv-m]{Chol-Jun Yu\corref{cor}}
\ead{ryongnam14@yahoo.com}
\cortext[cor]{Corresponding author}
\author[kimuniv-m]{Gum-Chol Ri}
\author[impp]{Andrew P. McMahon}
\author[impch]{Nicholas M. Harrison}
\author[impp]{Piers R. F. Barnes}
\author[imp]{Aron Walsh\corref{cor}}
\ead{a.walsh@imperial.ac.uk}
\address[kimuniv-m]{Department of Computational Materials Design, Faculty of Materials Science, Kim Il Sung University, \\ Ryongnam-Dong, Taesong District, Pyongyang, Democratic People's Republic of Korea}
\address[kimuniv-n]{Natural Science Centre, Kim Il Sung University, Ryongnam-Dong, Taesong District, Pyongyang, Democratic People's Republic of Korea}
\address[imp]{Department of Materials, Imperial College London, London SW7 2AZ, United Kingdom}
\address[impp]{Department of Physics, Imperial College London, London SW7 2AZ, United Kingdom}
\address[impch]{Department of Chemistry, Imperial College London, London SW7 2AZ, United Kingdom}

\begin{abstract}
The use of methylammonium (MA) lead halide perovskites \ce{CH3NH3PbX3} (X=I, Br, Cl) in perovskite solar cells (PSCs) has made great progress in performance efficiency during recent years. However, the rapid decomposition of \ce{MAPbI3} in humid environments hinders outdoor application of PSCs, and thus, a comprehensive understanding of the degradation mechanism is required. To do this, we investigate the effect of water intercalation and hydration of the decomposition and ion migration of \ce{CH3NH3PbX3} using first-principles calculations. We find that water interacts with \ce{PbX6} and MA through hydrogen bonding, and the former interaction enhances gradually, while the latter hardly changes when going from X=I to Br and to Cl. Thermodynamic calculations indicate that water exothermically intercalates into the perovskite, while the water intercalated and monohydrated compounds are stable with respect to decomposition. More importantly, the water intercalation greatly reduces the activation energies for vacancy-mediated ion migration, which become higher going from X=I to Br and to Cl. Our work indicates that hydration of halide perovskites must be avoided to prevent the degradation of PSCs upon moisture exposure.
\end{abstract}


\maketitle

\section{Introduction}
Perovskite solar cells (PSCs) using methylammonium lead halide perovskites \ce{MAPbX3} (MA = methylammonium cation \ce{CH3NH3+}; X = halogen anion \ce{I-}, \ce{Br-} or \ce{Cl-}) as light harvesters have opened up a new stage in the field of photovoltaics. 
The rapid rise of power conversion efficiency of PSCs from the initial 3.81\% reported by Kojima {\it et al.} in 2009~\cite{Kojima} to over 20\%~\cite{Jeon,Bi} within a few years is quite remarkable, as compared with other types of solar cells~\cite{Sum,Luo,Xiao}. Moreover, much low cost of cell fabrication by chemical methods and much abundance of raw materials have attracted great attention, inspiring people with hope that solar power will become competitive with fossil fuels in the electricity market in the near future. In spite of a great advance in research and development of PSCs during the past years, there still remain some obstacles to prevent commercialization of PSCs, such as toxicity of lead~\cite{Hailegnaw} and mostly poor material stability of these halide perovskites~\cite{NiuRev,Li,Wang,Berhe,Manser}.

First-principles calculations revealed that \ce{MAPbI3}, the most widely used hybrid perovskite halide in PSCs, may be decomposed exothermally~\cite{Zhang} and such intrinsic instability can be cured by mixing halogen ions~\cite{yucj10,yucj12}. 
On the other hand, moisture has been reported to play a critical role as an extrinsic factor in the degradation of PSC performance~\cite{Noh}. Although controlled humidity condition or a small amount of \ce{H2O} has positive effects on PSC performance such as enhancement of the reconstruction in perovskite film formation and triggering nucleation and crystallization of perovskite phase~\cite{Zhou14,Bass,You14}, \ce{H2O} molecule is known to damage the integrity of perovskite films. Some researchers asserted that \ce{MAPbI3} may readily hydrolyze in the presence of water due to deprotonation of \ce{CH3NH3+} by \ce{H2O}, resulting in degradation products such as \ce{CH3NH2}, \ce{HI} and \ce{PbI2}~\cite{NiuRev,Frost14,Niu}. 
On the contrary, different degradation reactions in perovskite films under humidity condition have been reported. Huang {\it et al.}~\cite{Huang} focused on the interaction between \ce{H2O} and \ce{MAPbI3} and suggested that degradation of \ce{MAPbI3} under ambient condition may produce \ce{PbCO3} and $\alpha$-PbO rather than \ce{PbI2}. 

Some reports found that the formation of monohydrated \ce{CH3NH3PbI3$\cdot$H2O} or dehydrated \ce{(CH3NH3)4PbI6$\cdot2$H2O} intermediate phase is the initial step in the perovskite decomposition process under 80\% humidity exposure~\cite{Yang,Hao,Christians,Leguy}. It was found subsequently that these intermediate phases are unstable in ambient condition, so that further decomposition into the final products could occur~\cite{Conings,Wozny,Bryant,Xiao16,Zhao,Ko}. As a consequence, the decomposition of \ce{MAPbI3} in humid air is a rather complicated process and their reaction processes or mechanisms are in an active controversy. To unveil the mechanism of \ce{MAPbI3} decomposition upon humidity exposure, theoretical simulations based on the density functional theory (DFT) have been performed, focused on \ce{H2O} adsorption on \ce{MAPbI3} surfaces~\cite{Tong,Mosconi15,Zhang15,Koocher}. However, comprehensive understanding of water-assisted decomposition of \ce{MAPbI3} is not yet fully established, and is urgent to facilitate materials engineering for enhanced material stability.

In this work, we investigate the influence of water intercalation and hydration on the decomposition of \ce{MAPbX3} (X = I, Br, Cl) by performing first-principles calculations. The crystalline and atomistic structures of water intercalated phases, as we denoted \ce{MAPbX3}\_\ce{H2O}, and monohydrated phases \ce{MAPbX3$\cdot$H2O} are explored carefully. These phases can be regarded as the intermediate ones on the process of water-assisted decomposition of \ce{MAPbX3}. The intercalation energies of a water molecule into the pseudo-cubic phases and the decomposition energies of the water intercalated and monohydrated phases into \ce{PbX2}, \ce{CH3NH3X} and \ce{H2O} are calculated to draw a meaningful conclusion about the stability. Finally we consider vacancy-mediated diffusion of an \ce{X-} anion, \ce{MA+} cation, and \ce{H2O} molecule in pristine, water intercalated and monohydrated phases, since such migrations seem to give important implication of the material stability~\cite{Eames,Haruyama,Egger,Azpiroz, Du}. \ce{Pb^2+} migration is excluded due to high formation energy of the Pb vacancy, \ce{V_{Pb}}.

\section{Computational Methods}

All of the pimary DFT calculations were carried out using the pseudopotential plane wave method as implemented in Quantum-ESPRESSO package~\cite{QE}. We used the ultrasoft pseudopotentials provided in the package\footnote{The pseudopotentials C.pbe-n-rrkjus\_psl.0.1.UPF, H.pbe-rrkjus\_psl.0.1.UPF, N.pbe-n-rrkjus\_psl.0.1.UPF, Pb.pbe-dn-rrkjus\_psl.0.2.2.UPF, and I(Br,Cl).pbe-n-rrkjus\_psl.0.2.UPF were used.}, where the valence electronic configurations of atoms are H--$1s^1$, C--$2s^22p^2$, N--$2s^22p^3$, Cl--$3s^23p^5$, Br--$4s^24p^5$, I--$5s^25p^5$, and Pb--$5d^{10}6s^26p^2$. The exchange-correlation interaction between the valence electrons was estimated using the Perdew-Burke-Ernzerhof (PBE)~\cite{pbe} form within the generalized gradient approximation, which is augmented by dispersive van der Waals interaction (vdW-DF-OB86), already shown to be important for calculations of perovskite halides~\cite{Geng,yucj08}. The structures of \ce{MAPbX3} and water intercalated phases were assumed to be pseudo-cubic. Unit cells containing one formula unit (f.u.) for these phases and two formula units for monohydrated phase were used for structural optimization and decomposition energetics. For diffusion process, (2$\times$2$\times$2) supercells for pseudo-cubic phase and (2$\times$2$\times$1) supercells for monohydrated phases, including 120 and 96 atoms, were used. The cutoff energy for plane-wave basis set is as high as 40 Ry and $k$-points are ($2\times2\times2$) with Monkhorst-Pack method, which guarantee the total energy accuracy as 5 meV per unit cell. Atomic positions were fully relaxed until the forces converge to $5\times10^{-5}$ Ry/Bohr. The activation energies for migrations were calculated using the climbing image nudged elastic band (NEB) method~\cite{NEB}.

\begin{figure}[!t]
\begin{center}
\footnotesize
\begin{tabular}{cc}
\includegraphics[clip=true,scale=0.17]{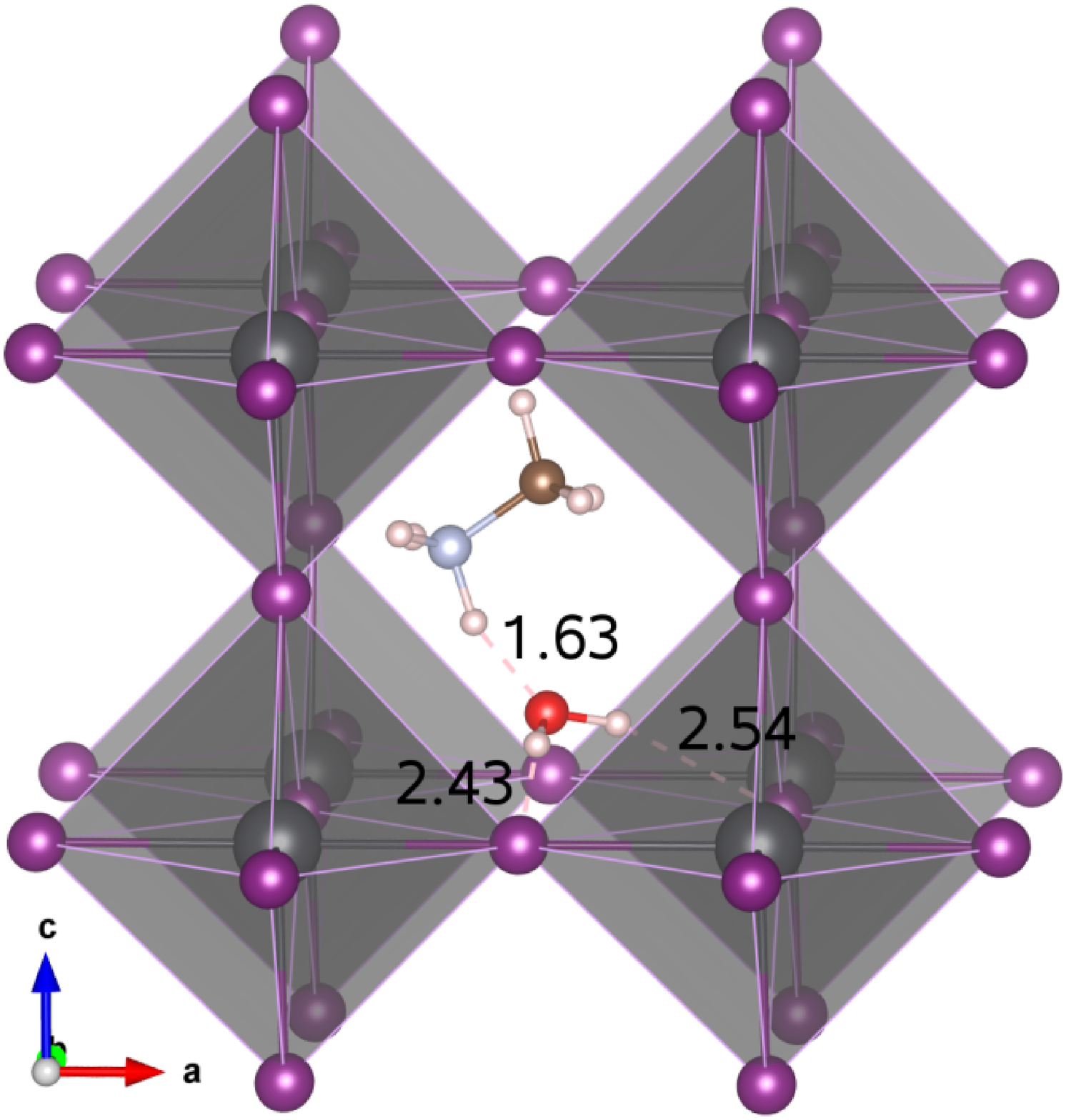} & \includegraphics[clip=true,scale=0.21]{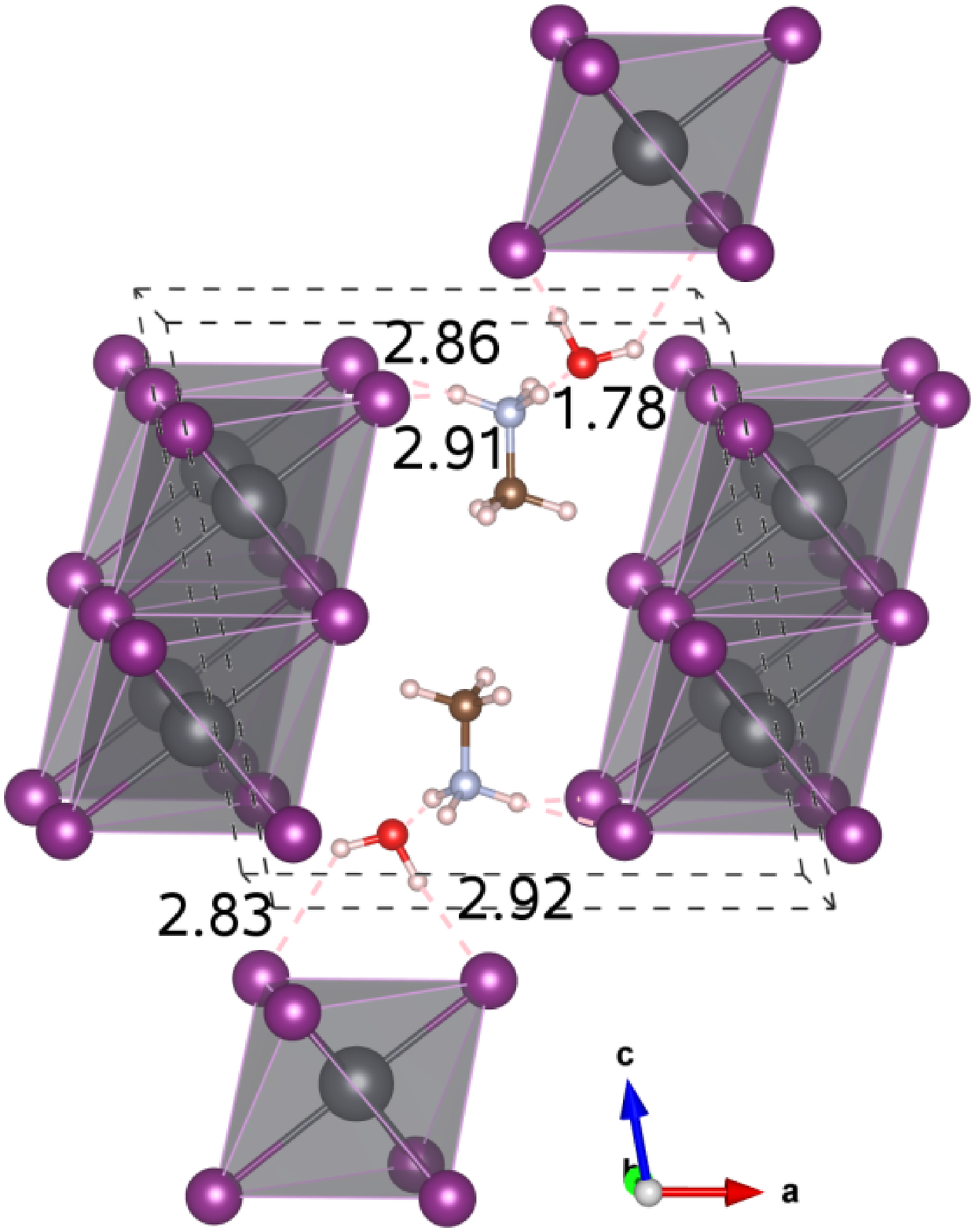}  \\
(a) & (b) \\ 
\end{tabular}
\end{center}
\caption{\label{fig_cell}Polyhedral view of water intercalated phase \ce{MAPbI3}\_\ce{H2O} (a) and monohydrated phase \ce{MAPbI3$\cdot$H2O} (b). Hydrogen bonds are marked with dotted lines and the bond lengths are presented in units of $\AA$ngstrom (dark grey: Pb; purple: I; brown: C; light blue: N; red: O; light pink: H).}
\end{figure}

The structural optimizations of pseudo-cubic \ce{MAPbX3} phases produced lattice constants of 6.330, 5.949, and 5.682 \AA~for X=I, Br, and Cl, which are in good agreement with the experimental values~\cite{Poglitsch}~within 1\% relative errors. A water molecule was put into the interstitial space of these optimized cubic phases, which were re-optimized. The supercell shape becomes triclinic following optimization. We have also performed optimization of the monohydrated phases \ce{MAPbX3$\cdot$H2O} with monoclinic crystalline lattice and experimentally identified atomic positions~\cite{Hao,Leguy,Imler,Vincent}. For the case of \ce{MAPbI3$\cdot$H2O}, the determined lattice constants $a=10.460$ \AA, $b=4.630$ \AA, $c=11.100$ \AA, and $\beta=101.50^\circ$ agree well with the experimental results~\cite{Imler}. It is worth noting that the \ce{MAPbI3} crystal has a 3-dimensional structure with corner-sharing \ce{PbI6} octahedra, while the monohydrated phase \ce{MAPbI3\cdot H2O} is characterized by a 1-dimensional edge-sharing \ce{PbI6} structure. The optimized atomistic structures of water intercalated phase \ce{MAPbI3}\_\ce{H2O} and monohydrated phase \ce{MAPbI3\cdot H2O} are shown in Figure~\ref{fig_cell}. 
It should be emphasized that the total energies per formula unit of \ce{MAPbX3}\_\ce{H2O} are typically 0.3 eV higher than those of \ce{MAPbX3\cdot H2O}, indicating that the water intercalated phase would be an intermediate phase in a transformation to the monohydrated phase.

In order to compare the free energies of formation for pristine and hydrated phases, additional calculations of the harmonic phonon density of states were calculated using the codes Phonopy\cite{phonopy} and VASP (PBEsol functional).\cite{vasp}
The Gibbs free energy of formation ($G_{solid}$) is calculated from the sum of the DFT internal energy ($U$), the vibrational free energy ($F_{vib}$), and the configurational entropy ($S_{conf}$).
\begin{equation}
G_{solid} = U + F_{vib} - TS_{conf}
\end{equation}
$S_{conf}$ takes account of the configurational entropy (rotational activity) of the MA$^+$ ion
from statistical mechanics, which 
cannot be extracted directly from static DFT calculations.\cite{lucy} 
To complement the solid-state calculations, the free energy of water vapour was estimated from 
a standard ideal gas expression. 

\section{Results}

\subsection{Water-perovskite interaction}

For the water intercalated phase \ce{MAPbI3}\_\ce{H2O}, the intercalated water molecule, with organic molecular \ce{MA+} ion, resides in the large interstitial space formed by a huge framework consisting of Pb and I atoms of inorganic \ce{PbI6} octahedra (for X=Br and Cl, similar structures are observed). It is bonded with H atoms of \ce{NH3} moiety of MA and I atoms of \ce{PbI6}; the bond length between the O atom of water and the H atom of \ce{NH3} is $\sim$1.63 \AA, and those between the H atoms of water and the I atoms of \ce{PbI6} are 2.43 and 2.54 \AA. In Figure~\ref{fig_cell}(b) for the monohydrated phase \ce{MAPbI3$\cdot$H2O}, the corresponding bond lengths are observed to be slightly longer, {\it i.e.}, 1.78, 2.83, and 2.92 \AA. These indicate that the interactions between water molecule and inorganic \ce{PbI6} as well as organic MA are through hydrogen bonding, as already pointed out in the previous works~\cite{Zhang15,Quarti,Gottesman}. 

When going from I to Br and to Cl, the (\ce{H2O})O$-$H(MA) bond length increases a little for the water intercalated phases or hardly change for the monohydrated phases, whereas the (\ce{H2O})H$-$X(\ce{PbX6}) bond lengths decrease distinctly (see Table~\ref{tab_bond}). 
On consideration that water intercalation causes a volume expansion, we estimate a relatively volume expansion rate as $r_\text{vol}=(V-V_0)/V_0\times100$\%, where $V$ is the volume of water intercalated or monohydrated phase and $V_0$ the volume of pseudo-cubic \ce{MAPbX3} phase per formula unit. This gradually increases going from X=I to Br and to Cl. Consequently, it can be deduced that decreasing the atomic number of halogen component induces an enhancement of the interaction between water and inorganic \ce{PbX6} matrix through hydrogen bonding, while maintaining those interactions between water and the MA ion, resulting in the contraction of the interstitial space and the volume, which might cause difficulty of water intercalation and ion migration.
\begin{table}[!t]
\caption{\label{tab_bond}Hydrogen bond length (\AA) and relative volume expansion rate of water intercalated phase \ce{MAPbX3}\_\ce{H2O} and monohydrated phase \ce{MAPbX3$\cdot$H2O}.}
\small
\begin{center}
\begin{tabular}{llcccc}
\hline 
& X & O-H(MA) & H-X1 & H-X2 & $r_\text{vol}$ (\%)\\ 
\hline
\ce{MAPbX3}\_\ce{H2O}      & I  & 1.63 & 2.43 & 2.54 & 7.10 \\
                           & Br & 1.66 & 2.24 & 2.31 & 10.18 \\
                           & Cl & 1.69 & 2.09 & 2.15 & 13.39 \\ \hline
\ce{MAPbX3}$\cdot$\ce{H2O} & I  & 1.78 & 2.83 & 2.92 & 4.40 \\
                           & Br & 1.77 & 2.65 & 2.81 & 9.40\\
                           & Cl & 1.79 & 2.34 & 2.83 & 10.21\\
\hline
\end{tabular}
\end{center}
\normalsize
\end{table}

To estimate the ease of water intercalation, we calculated the intercalation energy by
\begin{equation}
 E_\text{int}=E_{\ce{MAPbX3}\cdot\ce{H2O}}-(E_{\ce{MAPbX3}}+E_{\ce{H2O}})
\end{equation}
where $E_{\ce{MAPbX3}\cdot\ce{H2O}}$, $E_{\ce{MAPbX3}}$, and $E_{\ce{H2O}}$ are the total energies of the water intercalated or monohydrated bulk, pseudo-cubic bulk unit cells per formula unit, and an isolated water molecule, respectively. The intercalation energies in the water intercalated phases were calculated to be $-$0.53, $-$0.43, and $-$0.36 eV for X=I, Br, and Cl, which are smaller in magnitude than those in the monohydrated phases as $-$0.87, $-$0.78, and $-$0.69 eV, as presented in Table~\ref{tab_ene}. Therefore, it can be said that formation of the monohydrated phases is easier than formation of water intercalated phases. When going from X=I to Br and to Cl, the magnitude of water intercalation energy decreases, indicating that water intercalation becomes more difficult as the atomic number of halogen component decreases. 
This can be interpreted with the analysis of the interaction between water and perovskite, and the structural property. It should be noted that, although negative intercalation energies imply the exothermic process of water intercalation, a certain amount of energy (kinetic barrier) could be required for water molecule to intercalate into the perovskite halides, as penetration of water into the perovskite iodide surface~\cite{Tong,Mosconi15,Zhang15,Koocher}.
\begin{table}[!t]
\begin{center}
\caption{\label{tab_ene}Intercalation energy ($E_\text{int}$) of water into \ce{MAPbX3} perovskite halides and decomposition energy ($E_\text{dec}$) of water intercalated phase \ce{MAPbX3}\_\ce{H2O} and monohydrated phase \ce{MAPbX3$\cdot$H2O} (unit: eV).}
\small
\begin{tabular}{llcc}
\hline 
 & X & $E_\text{int}$ & $E_\text{dec}$ \\ 
\hline
\ce{MAPbX3}\_\ce{H2O}      & I  & $-$0.53 & 0.47 \\
                           & Br & $-$0.43 & 0.54 \\
                           & Cl & $-$0.36 & 0.60 \\ \hline
\ce{MAPbX3}$\cdot$\ce{H2O} & I  & $-$0.87 & 0.81 \\
                           & Br & $-$0.78 & 0.88 \\
                           & Cl & $-$0.69 & 0.92 \\
\hline
\end{tabular}
\end{center}
\end{table}

We further considered decomposition of water intercalated and monohydrated phases by calculating the decomposition energy,
\begin{equation}
 E_\text{dec}=(E_{\ce{PbX2}}+E_{\ce{MAX}}+E_{\ce{H2O}})-E_{\ce{MAPbX3}\cdot\ce{H2O}}
\end{equation}
where $E_{\ce{PbX2}}$ and $E_{\ce{MAX}}$ are the total energies of crystalline \ce{PbX2} (space group: $P$\={3}$m1$) and \ce{MAX} (space group: $Fm$\={3}$m$), respectively. As shown in Table~\ref{tab_ene}, the decomposition energies were calculated to be positive, indicating that the decomposition is an endothermic process. At this moment, it is worth to compare with those for the pristine perovskite halides without water. The decomposition of \ce{MAPbI3} $\rightarrow$ \ce{MAI} + \ce{PbI2} is exothermic due to the negative decomposition energy of $-$0.06 eV, whereas those of \ce{MAPbBr3} and \ce{MAPbCl3} are also endothermic with the positive decomposition energies of 0.10 and 0.23 eV~\cite{yucj10,yucj12}. Interestingly, the decomposition energy in magnitude is in the reverse order to the intercalation energy going from X=I to Br and to Cl. This indicates that water could intercalate more readily into the halide perovskite but the formed water-included compounds would be more resistant to decomposition into their relevant components as decreasing the atomic number of halogen component.

\begin{figure}[!t]
\begin{center}
\footnotesize
\begin{tabular}{cc}
\includegraphics[]{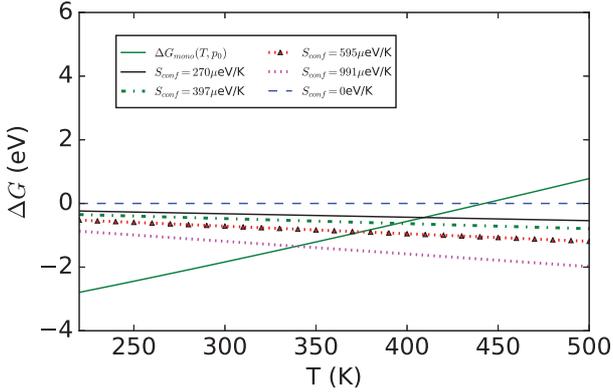} \\
\end{tabular}
\end{center}
\caption{\label{figx} Calculated Gibbs free energies as a function of temperature for the monohydrate at fixed pressure $p_0$ = 1 bar. Also shown are the reference free energies of MAPbI$_3$ assuming different values for the (statistical mechanical) configurational entropy due to the motion of the MA$^+$ cation, $S_{conf}$. Additional entropy shifts the transition temperature for hydrate formation to lower temperature values (from T=445 K to T=345 K). }
\end{figure}

To consider finite-temperature effects, additional calculations were performed to compute the Gibbs free energy of each phase (see  Figure \ref{figx}). 
Formation of the hydrated compound is found to be favourable at lower temperatures due to the large energy gain from creation of the hydrate versus a state containing MAPbI$_3$ and water vapour.
The calculations confirm the results stated earlier, but also predict that at higher temperatures 
the pristine MAPI$_3$ phase becomes stabilized by its vibrational entropy. 
The inclusion of configuration entropy associated with the rotational freedom of the MA$^+$ ion acts to further stabilize the material against hydration.

\subsection{Dynamics of water incorporation and charged defects}

We turn our attention to how the ions and water molecule diffuse inside the water intercalated or monohydrated perovskites, trying to find out their role in material instability. It is well known that diffusion of ions in crystalline solid is associated with point defects such as a site vacancy and/or interstitial. While such migrations of ions or defects can provide explicit explanations for the performance of PSC device such as ionic conduction, hysteresis, and field-switchable photovoltaic effect~\cite{Eames,Haruyama,Egger,Azpiroz,Du,Frost,Yin,Walsh}, these might have important implications for material stability.

 It was established that, although as in other inorganic perovskite oxides several types of point defects could be formed in the hybrid perovskite halides, including vacancies (\ce{V_{MA}}, \ce{V_{Pb}}, \ce{V_X}), interstitials (\ce{MA_i}, \ce{Pb_i}, \ce{X_i}), cation substitutions (\ce{MA_{Pb}}, \ce{Pb_{MA}}) and antisite substitutions (\ce{MA_X}, \ce{Pb_X}, \ce{X_{MA}}, \ce{X_{Pb}}), vacancies except \ce{V_{Pb}} have the lowest formation energies, while others are unstable both energetically and kinetically~\cite{Du,Walsh}. In this work we thus considered only vacancies \ce{V_{MA}} and \ce{V_X} in the pristine, water-intercalated, and monohydrated phases, which could support vacancy-mediated ionic diffusion. For migration of the water molecule inside the water intercalated or monohydrated phases, water vacancy \ce{V_{H$_2$O}} is formed and allowed to migrate.

\begin{figure}[!t]
\begin{center}
\includegraphics[clip=true,scale=0.2]{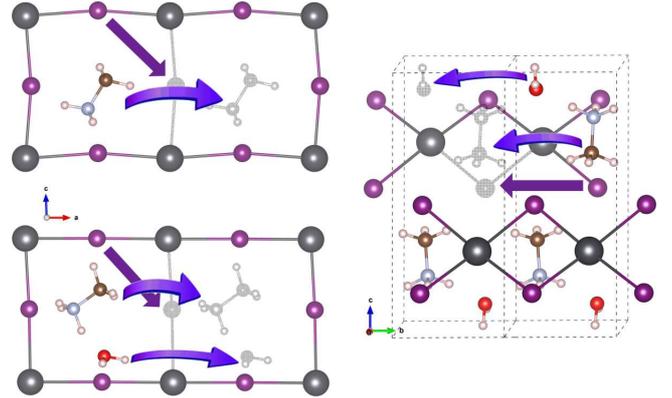}
\end{center}
\caption{\label{fig_migscheme}Schematic view of vacancy-mediated ion and molecule migrations in the pristine (left-top), water-intercalated (left-bottom), and monohydrated (right) perovskite halides. Colour scheme for atoms is the same to Figure~\ref{fig_cell}.}
\end{figure}

For vacancy-mediated ion migration in the pristine and water intercalated \ce{MAPbX3} phases, we follow the three vacancy transport mechanisms established in the previous works~\cite{Eames,Haruyama,Azpiroz}, where vacancies are allowed to conventionally hop between neighbouring equivalent sites. According to these mechanisms, \ce{X-} at a corner site of \ce{PbX6} octahedron migrates along the octahedron edge towards a vacancy in another corner site, and \ce{MA+} hops into a neighbouring vacant cage formed by the inorganic scaffold. Water molecule migrates along the similar path to \ce{MA+} case. For the cases of monohydrated phase, we have devised plausible paths for each of the three defects, and pick out one that has the lowest activation energy, as discussed below in detail. Figure~\ref{fig_migscheme} shows the schematic view of vacancy-mediated ion and molecule migration paths. Special attention was paid to obtaining the well-converged structures of the start and end point configurations with structural relaxations with convergence criteria as 0.01 eV/\AA~atomic forces. The activation energies for these vacancy-mediated migrations are summarised in Table~\ref{tab_neb}.
\begin{table}[!b]
\caption{\label{tab_neb}Activation energies (eV) of vacancy-mediated \ce{X-}, \ce{MA+} ions, and \ce{H2O} molecule migrations in the pristine pseudo-cubic phases \ce{MAPbX3}, water-intercalated phases \ce{MAPbX3}\_\ce{H2O}, and monohydrated phases \ce{MAPbX3$\cdot$H2O}.}
\begin{center}
\begin{tabular}{llccc}
\hline 
  & X & \ce{X-} & \ce{MA+} & \ce{H2O} \\ 
\hline
\ce{MAPbX3}               & I  & 0.55 & 1.18 &      \\
                          & Br & 0.58 & 1.20 &      \\
                          & Cl & 0.62 & 1.24 &      \\ \hline
\ce{MAPbX3}\_\ce{H2O}     & I  & 0.22 & 0.38 & 0.28 \\
                          & Br & 0.29 & 0.54 & 0.31 \\
                          & Cl & 0.35 & 0.63 & 0.42 \\ \hline
\ce{MAPbX3}$\cdot$\ce{H2O}& I  & 0.44 & 1.14 & 0.78 \\
                          & Br & 0.47 & 1.18 & 0.89 \\
                          & Cl & 0.49 & 1.23 & 1.08 \\
\hline
\end{tabular}
\end{center}
\end{table}

To see whether our computational models and parameters could give reasonable results for the ionic migrations, the pseudo-cubic \ce{MAPbI3} was first tested. As listed in Table~\ref{tab_neb}, the activation energies for \ce{I-} and \ce{MA+} migrations were calculated to be 0.55 and 1.18 eV, respectively, which are comparable with 0.58 and 0.84 eV reported in ref.~\cite{Eames}, and 0.32$-$0.45 and 0.55$-$0.89 eV in ref.~\cite{Haruyama}, but higher than 0.16 and 0.46 eV in ref.~\cite{Azpiroz}. With respect to the crystalline lattice, we used the pseudo-cubic lattice as in the work in ref.~\cite{Eames}, while Haruyama {\it et al.}~\cite{Haruyama} and Azpiroz {\it et al.}~\cite{Azpiroz} used the tetragonal lattice. An exchange-correlation (XC) functional including dispersion (vdW) interactions was used in our work and the work in ref.~\cite{Haruyama}, whereas PBEsol and PBE without vdW correction were used in ref.~\cite{Eames} and ref.~\cite{Azpiroz}, respectively. Therefore, the slight discrepancies might be associated with the different crystalline lattices and XC functionals without vdW correction, not with the supercell size. Most important, the activation energy for \ce{I-} migration is lower than that for \ce{MA+} migration in all the above-mentioned works, convincing that the results obtained in this work can be used to find out the influence of water on ion diffusion.

\begin{figure}[!t]
\begin{center}
\footnotesize
\begin{tabular}{cc}
 \includegraphics[clip=true,scale=0.15]{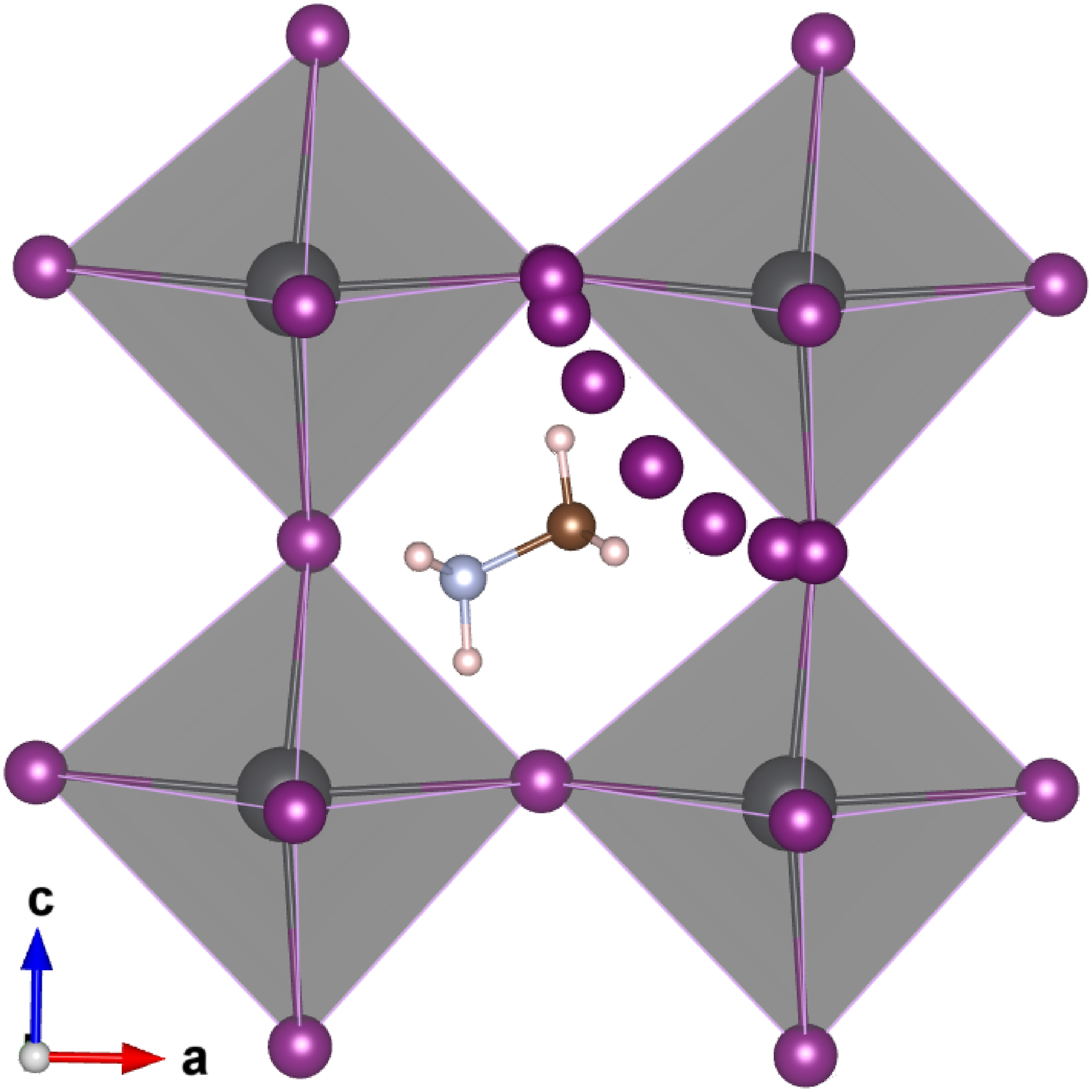} & 
 \includegraphics[clip=true,scale=0.15]{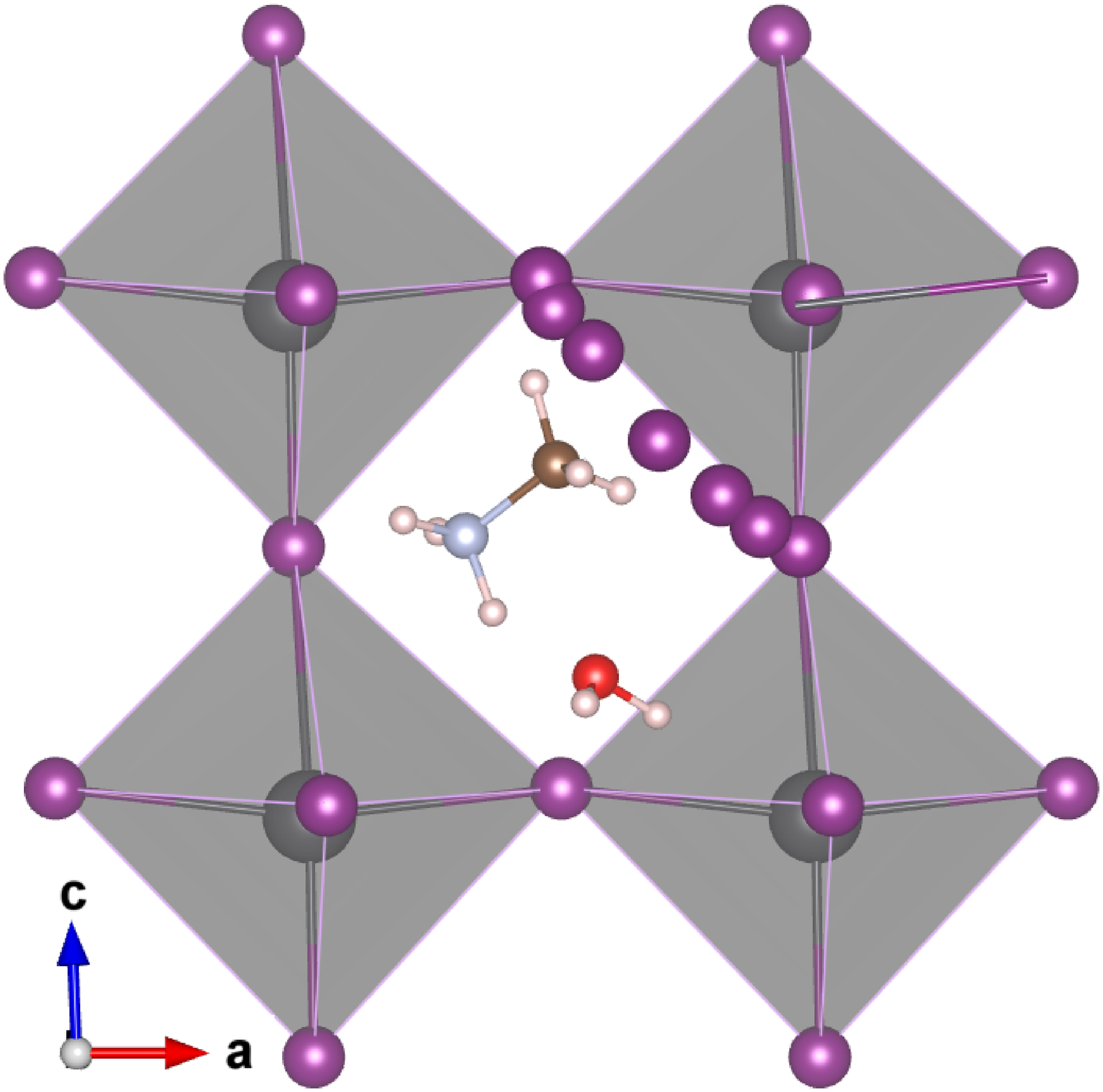}\\
 (a) & (b) \\ 
 \includegraphics[clip=true,scale=0.17]{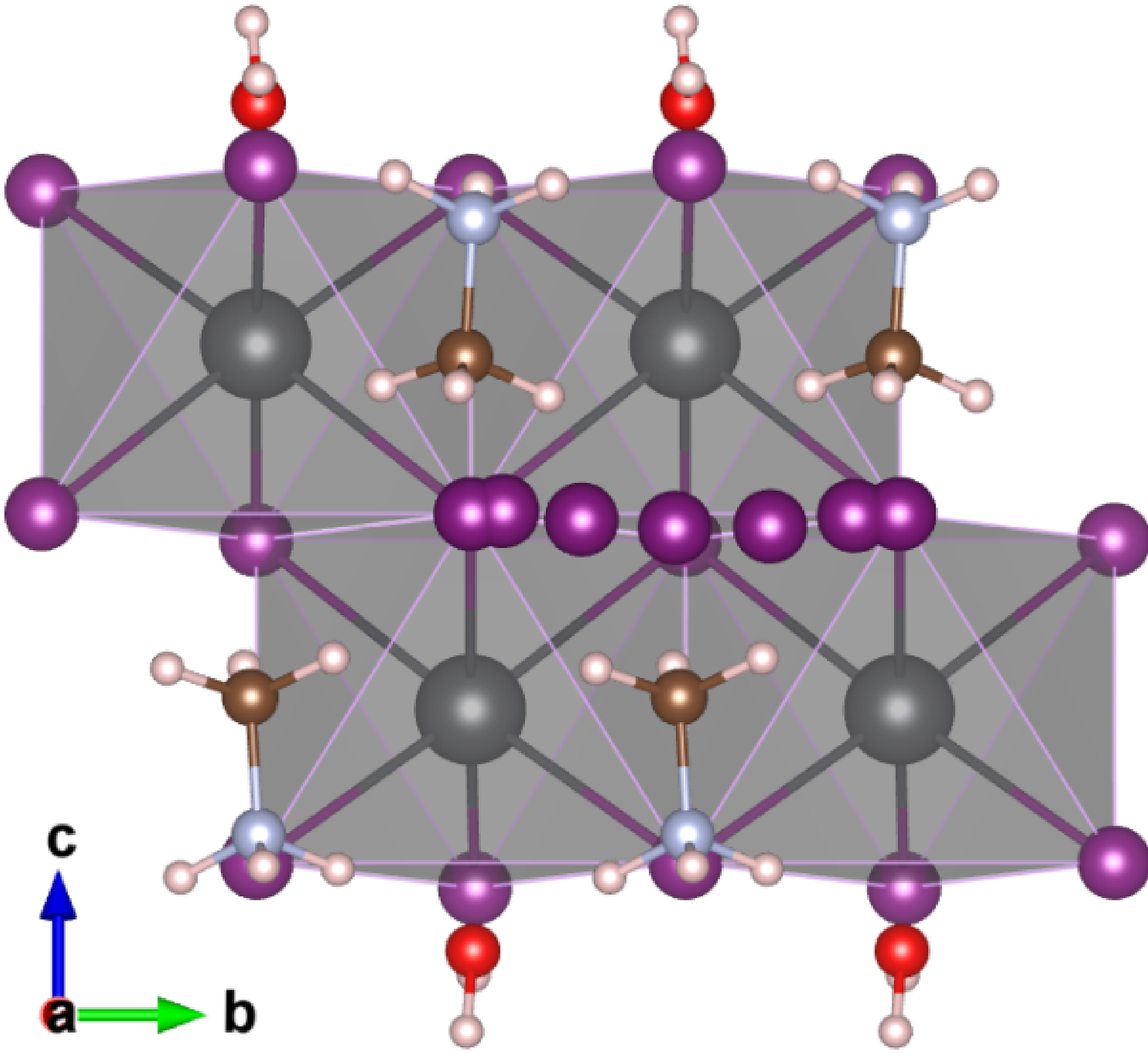} & 
 \includegraphics[clip=true,scale=0.37]{fig3d.eps} \\ 
 (c) & (d) \\ 
\end{tabular}
\end{center}
\caption{\label{fig_Xmig}\ce{I-} ion migrations in (a) pseudo-cubic \ce{MAPbI3}, (b) water intercalated phase \ce{MAPbI3\_H2O} and (c) monohydrated phase \ce{MAPbI3$\cdot$H2O}, and (d) corresponding activation energy profile. All the atoms are allowed to relax during migration, resulting in slight distortion of \ce{PbI6} octahedra.}
\end{figure}

For \ce{I-} migration in the monohydrated phase \ce{MAPbI3\cdot H2O}, which is structurally characterized by 1-dimensional edge sharing \ce{PbI6} octahedra connected in [010] direction, we devised four migration pathways along the three octahedron edges in different directions and across the space between separated octahedra. The lowest activation energy was found for the migration along the edge in [010] direction, while the highest value over 2 eV was found for the one across the space, implying this pathway less likely. Figure~\ref{fig_Xmig} shows the \ce{I-} migration pathways along the octahedron edge in the three kinds of phases and the corresponding activation energy profile. It is found that, when water intercalates into the perovskite, the activation energy decreases, indicating more facile diffusion of \ce{I-} ion upon water intercalation. Meanwhile, the activation energy in the monohydrated phase is higher than in the water intercalated phase but still lower than in the pristine phase. This demonstrates that the intercalated water molecule enhances diffusion of ions in hybrid perovskite halides, facilitating the formation of hydrated phases. However, once the hydrated phase is formed, diffusion of ions becomes a little harder. Similar arguments hold for \ce{MAPbBr3} and \ce{MAPbCl3}.

\begin{figure}[!t]
\begin{center}
\footnotesize
\begin{tabular}{cc}
 \includegraphics[clip=true,scale=0.11]{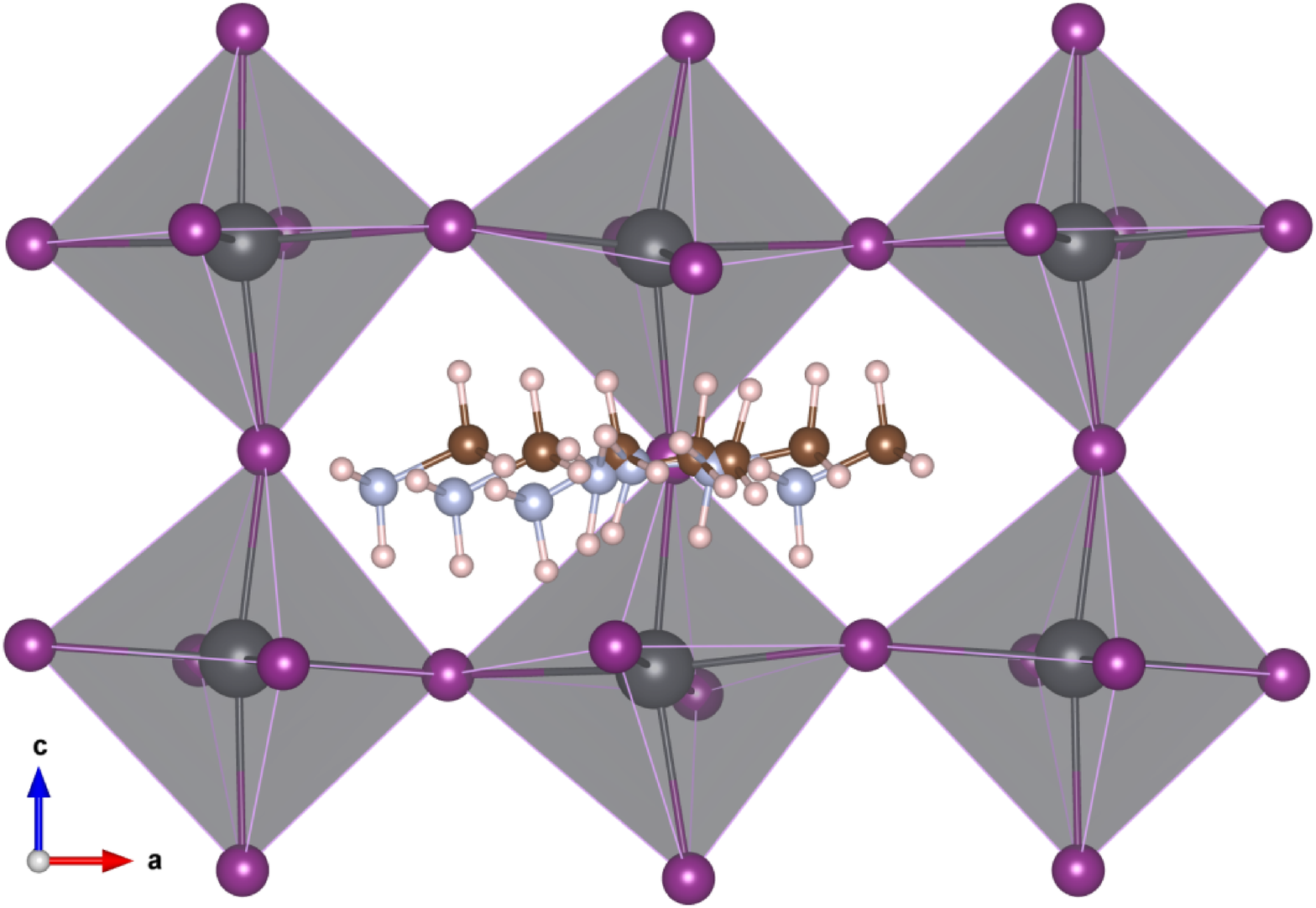} & 
 \includegraphics[clip=true,scale=0.11]{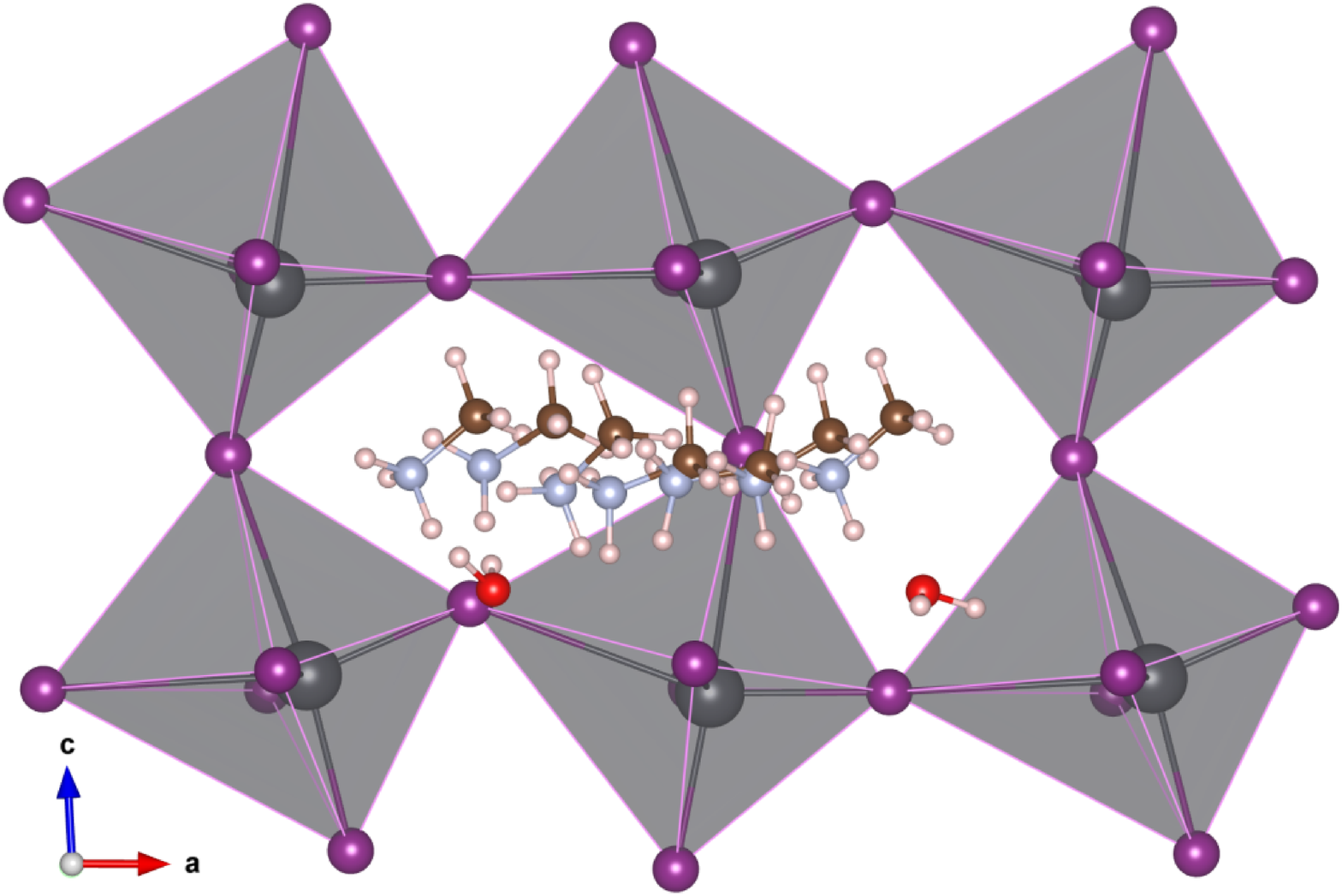}\\ 
 (a) & (b) \\ 
 \includegraphics[clip=true,scale=0.13]{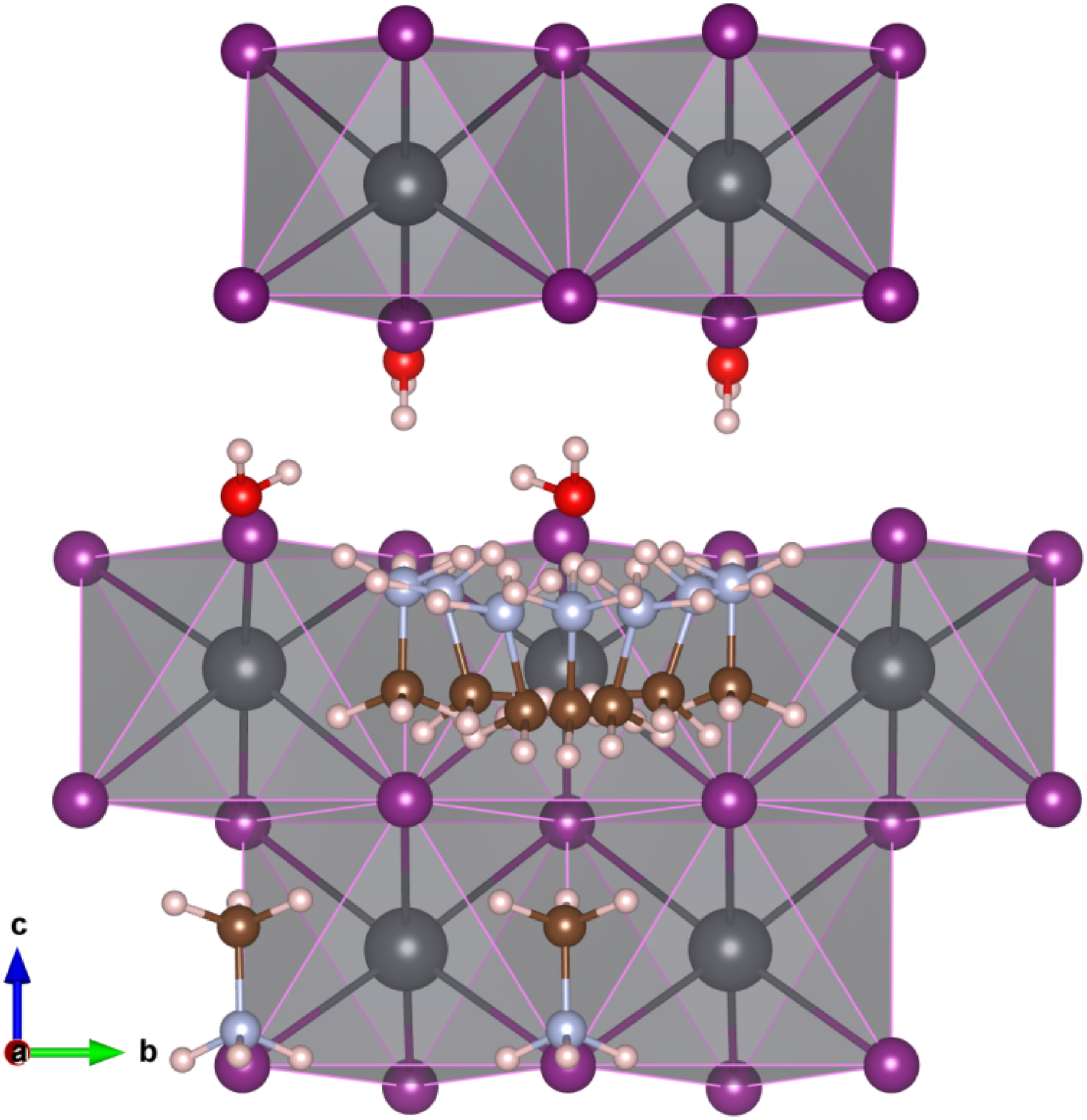} & 
 \includegraphics[clip=true,scale=0.37]{fig4d.eps} \\
 (c) & (d) \\ 
\end{tabular}
\end{center}
\caption{\label{fig_MAmig}\ce{MA+} ion migration in (a) pseudo-cubic \ce{MAPbI3}, (b) water intercalated phase \ce{MAPbI3\_H2O} and (c) monohydrated phase \ce{MAPbI3$\cdot$H2O}, and (d) corresponding activation energy profile.}
\end{figure}

An \ce{MA+} ion in the monohydrated phase was enforced to migrate along the almost straight pathway in [010] direction since there is no channel in [100] direction due to the wall formed by \ce{PbI6} octahedra and the channel in [001] direction has much longer distance. It should be noted that \ce{MA+} cation in the water intercalated phase is allowed to diffuse equally both in [100] and [010] directions but not allowed to move in [001] direction due to the presence of water molecule on the path. Figure~\ref{fig_MAmig} shows the intermediate states during \ce{MA+} migrations in the pristine, water-intercalated, and monohydrated \ce{MAPbI3} phases and the corresponding energy profile. We can see distortions of \ce{PbI6} octahedra during migration, much more clearly in the case of the water intercalated phase due to the rather strong interaction between \ce{PbI6} and methylammonium, which may cause difficult \ce{MA+} ion migration. 

Relatively high activation energies for \ce{MA+} ion migrations were found for the pristine (1.18 eV) and monohydrated phases (1.14 eV), while low activation energy of 0.38 eV was found for the water intercalated phase. This indicates that inclusion of water in the perovskite halides reduces the activation energy for \ce{MA+} ion migration as in the case of halogen ion migration. As discussed above, the volume expansion rate of the water intercalated phase is larger than that of the monohydrated phase, and thus, water intercalation can induce much more space expansion, resulting in the enhancement of \ce{MA+} ion diffusion. When compared to \ce{I-} migration, \ce{MA+} migration can be said to be more difficult due to higher activation energy in agreement with the previous works~\cite{Eames,Haruyama,Azpiroz}, in which the bottleneck comprising four \ce{I-} ions and the high level of orientational motion of the \ce{MA+} ion were pointed out to be the reasons for such hard migration.

\begin{figure}[!t]
\begin{center}
\footnotesize
\begin{tabular}{cc}
 \includegraphics[clip=true,scale=0.15]{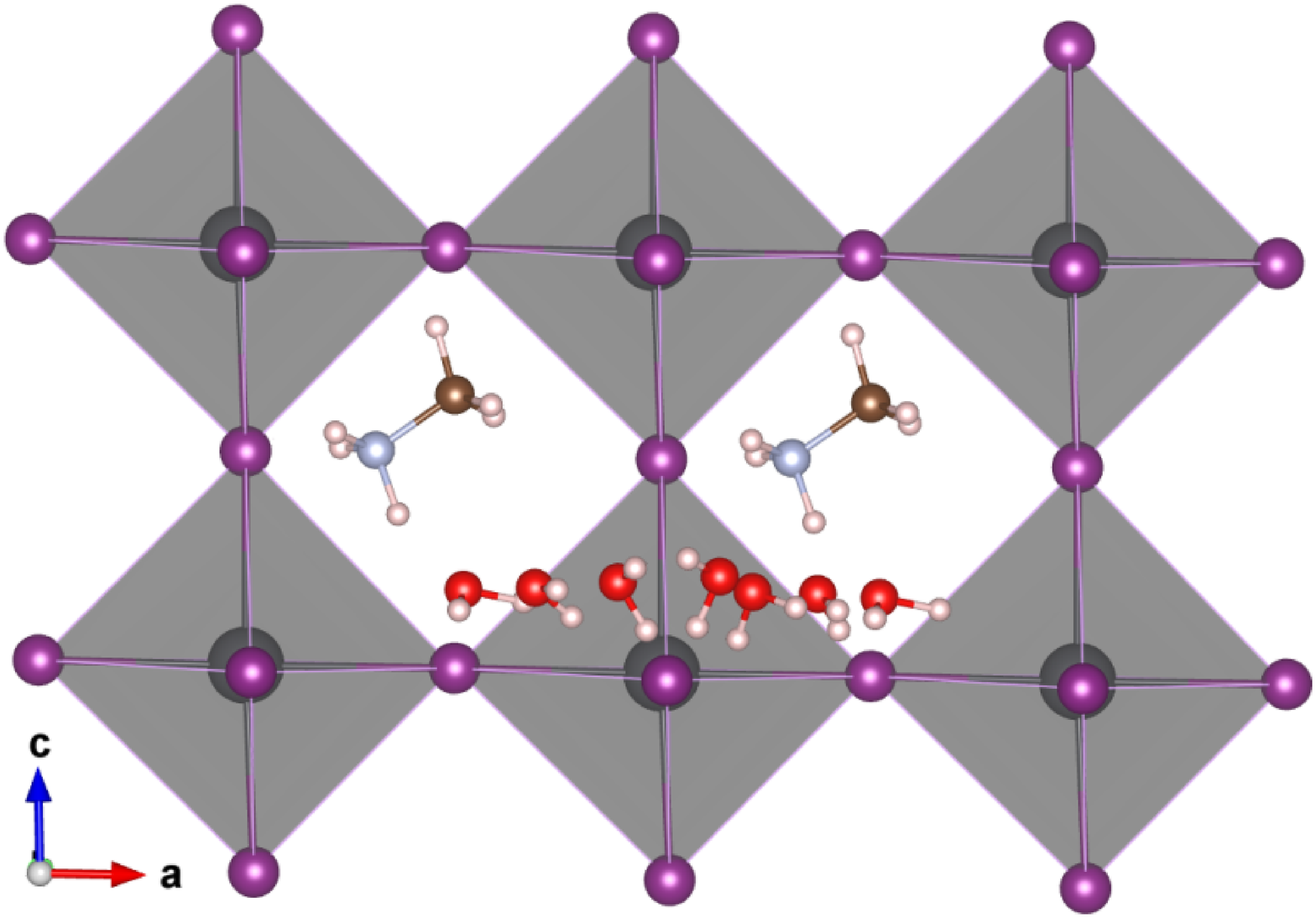} & 
 \includegraphics[clip=true,scale=0.13]{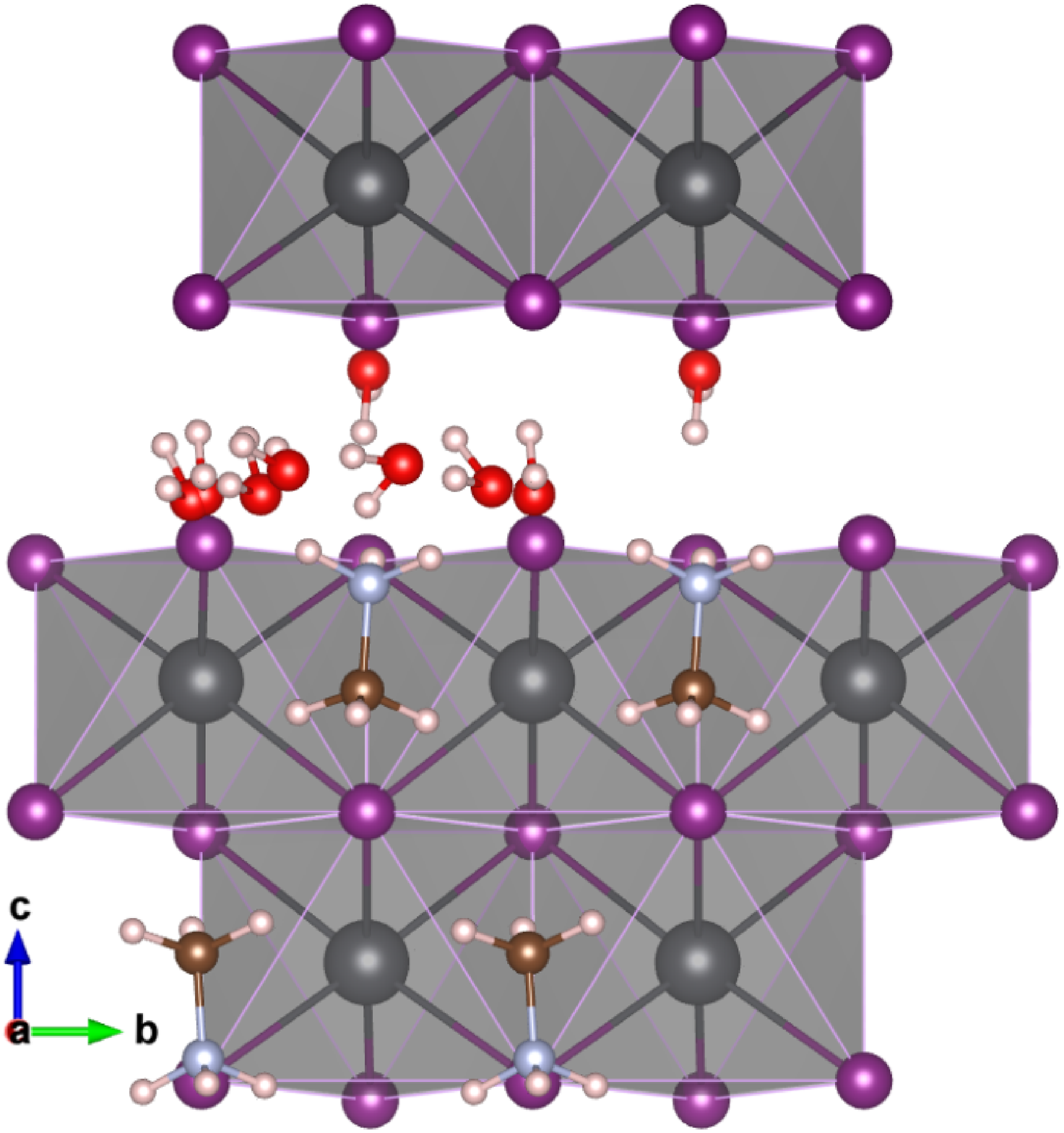}\\ 
 (a) & (b) \\ 
 \multicolumn{2}{c}{\includegraphics[clip=true,scale=0.38]{fig5c.eps}} \\
 \multicolumn{2}{c}{(c)} 
\end{tabular}
\end{center}
\caption{\label{fig_H2Omig}\ce{H2O} molecule migrations in (a) water intercalated phase \ce{MAPbI3\_H2O} and (b) monohydrated phase \ce{MAPbI3$\cdot$H2O}, and (c) corresponding activation energy profile.}
\end{figure}

Finally, a \ce{H2O} molecule was allowed to migrate in the same direction as \ce{MA+} ion in the water intercalated and monohydrated phases, as represented in Figure~\ref{fig_H2Omig}. The activation energy in the water intercalated phase was calculated to be 0.28 eV, which is low enough to diffuse inside the bulk crystal and form the hydrated phase. As in the cases of ion migrations, this is lower than the one in the monohydrated phase. At this stage, it is worth to compare with the initial process of water penetration into \ce{MAPbI3} surface. From the first-principles calculations~\cite{Koocher,Mosconi,Tong}, it was found that, when water molecules are brought into contact with \ce{MAPbI3} surface, the water molecule in the inside region is 0.2$\sim$0.3 eV more stable than the one in the outside region, and thus, the water molecule is strongly driven to diffuse into the inside of \ce{MAPbI3}. The activation barrier for this process was calculated to be 0.31 eV~\cite{Tong} or 0.27 eV~\cite{Koocher} at low coverage of water and 0.82 eV~\cite{Tong} at high coverage. These are comparable to the barrier of water diffusion within the bulk crystal in this work, which can be regarded as a continuation of the water penetration, indicating easy formation of water intercalated and further hydrated phase. As can be seen in Table~\ref{tab_neb}, water molecule can migrate more easily than \ce{MA+} ion, which might be due to larger molecular size of MA and its stronger interaction with environmental components, but more toughly than \ce{X-} ion in both phases. On the other hand, the activation barrier of water migration in the monohydrated phase is higher than the one in the water intercalated phase, as in the cases of \ce{X-} and \ce{MA+} ion migrations.

When the atomic number of halogen component decreases from X=I to Br and to Cl, the activation barriers for ions and water molecule migrations increases monotonically, as shown in Table~\ref{tab_neb}. 
In fact, when changing from I with larger ionic radius of 2.2 \AA~to Br with smaller ionic radius of 1.96 \AA~and to Cl with further smaller ionic radius of 1.81 \AA, the lattice spacing and \ce{PbX6}$-$MA bonding shrink while maintaining the intramolecular MA spacing~\cite{yucj10,yucj12}. This results in the enhancement of Pb$-$X interaction and makes passages of ions and water molecule more difficult. Similar arguments hold for water intercalated and monohydrated phases. The increase of activation barrier for ion migrations going from X=I to Cl describes enhancement of material stability when mixed I atom with Br or Cl atom, and is coincident with aforementioned decomposition energetics.

\section{Conclusion}
To understand the degradation mechanism of PSCs upon exposure of \ce{MAPbI3} to moisture, we have investigated the influence of water intercalation and hydration on decomposition and ion migrations of \ce{MAPbX3} (X=I, Br, Cl) by first-principles calculations. The crystalline lattices and atomistic structures of water intercalated \ce{MAPbX3} phases were suggested and optimized, together with those of monohydrated phases \ce{MAPbX3$\cdot$H2O}. It was found that water interacts with \ce{PbX6} and MA in these compounds through hydrogen bond, and the former interaction enhances gradually while the latter hardly change when going from X=I to Br and to Cl. The calculated results for water intercalation energies and decomposition energies indicate that water can exothermically intercalate into the hybrid perovskite halides, while the water intercalated and monohydrated compounds are stable with respect to decomposition. More importantly, the hydrogen bond interaction induced by water greatly affects the vacancy-mediated ion migrations, for which the activation barrier decreases upon the water inclusion inside the perovskite halides. The activation energies for ion and water molecule migrations become higher as going from X=I to Br and to Cl. These results clarify that degradation of the PSCs upon moisture exposure originates from the formation of water intercalated and further hydrated compounds and then decomposition of these compounds, which provides the idea to prevent this degradation at the atomic level.

\section*{Acknowledgments}
This work was supported partially by the State Committee of Science and Technology, Democratic People's Republic of Korea, under the state project ``Design of Innovative Functional Materials for Energy and Environmental Application'' (No. 2016-20). The work in the UK was supported by the Royal Society and the Leverhulme Trust, and the Imperial College High Performance Computing Service.
A.P.M. was supported by a studentship from the Centre for Doctoral Training in Theory and Simulation of Materials at Imperial College London, funded by the EPSRC under grant EP/G036888. 
The calculations have been carried out on the HP Blade System C7000 (HP BL460c) that is owned and managed by Faculty of Materials Science, Kim Il Sung University.

\section*{\label{note}Notes}
The authors declare no competing financial interest.

\bibliographystyle{rsc}
\bibliography{Reference}

\end{document}